\documentclass[prb,showpacs]{revtex4}
\usepackage{graphicx}
\begin{document}
\title{Relaxation of Femtosecond Non-equilibrium Electrons in a Metallic Sample} 
\author{Navinder Singh}
\email{navinder@iopb.res.in}
\affiliation{Institute of Physics, Bhubaneswar-751005, India}
\begin{abstract}
A model calculation is given for the energy relaxation of a non-equilibrium distribution of hot electrons prepared in a metallic sample that has been subjected to homogeneous photo-excitation by a femtosecond 
laser pulse. The model assumes that the delta pulse photoexcitation creates two interpenetrating electronic subsystems, initially comprising a  dilute energy-wise higher-lying non-degenerate hot electron subsystem, and a relatively dense, lower-lying electron subsystem which is degenerate. In  the femtosecond time regime the relaxation process is taken to be dominated by the electron-(multi) phonon  interaction, resulting in a quasi-continuous  electron energy loss to the phonon bath. The kinetic model is given for this time regime, beacuse in this time regime the usual Two Temperature model is not applicable. The Two Temperature model assumes that the hot electrons and phonons are in their respective equilibrium states (Fermi and Bose) but at a different temperatures. One uses the Bloch-Boltzmann-Peierls transport formula to calculate the energy transfer rate. 
In the present Kinetic model a novel physical feature of slowing down (due to Fermionic blocking of interaction phase space) of electron-phonon relaxation mechanism near the Fermi energy of degenerate electronic subsystem is considered. This leads to a peaking of the calculated hot electron distribution at the Fermi energy. This feature, as well as the entire evolution of the hot electron distribution, may be time-resolved by 
a femto-second pump-probe study.
\end{abstract}
\maketitle
\section{Introduction}
Consider the kinetic evolution of a photoexcited nonequilibrium 
electron-phonon system, comprising three subsystems. The first two are interpenetrating electronic subsystems, one degenerate and the other nondegenerate, while the third is the phonon bath (the lattice). For these three interacting subsystems, one can define three characteristic relaxation times, namely, the electron--phonon $(\tau_{e–p})$, the phonon--phonon $(\tau_{p–p})$, and the electron--electron $(\tau_{e–e})$ relaxation times. Thus, for example, in the case of metals, one usually has $\tau_{e–e}\ll\tau_{e–p}\gg\tau_{e–p}$. In fact, the electron--electron relaxation time is approximately one-tenth of that of the electron--phonon, the latter being a few picoseconds. For this very reason, one uses the standard Two Temperature model\cite{ani,kag,sch,sun,all}. The present model can be used for the initial femto-second time scales where the electron distribution is still  in non-equilibrium (non Fermi-Dirac Distribution) state.
\section{the model}
Consider a delta pulse photoexcitation creates two interpenetrating electronic subsystems, initially comprising a  dilute energy-wise higher-lying non-degenerate hot electron subsystem, and a relatively dense, lower-lying electron subsystem which is degenerate. In  the femtosecond time regime the relaxation process is taken to be dominated by the electron-(multi)phonon  interaction, resulting in a quasi-continuous  electron energy loss to the phonon bath. The femtosecond photoexcited non-degenerate hot electron distribution then evolves predominately through the non-radiative
electron-phonon processes. In the present work, two models of electron-(multi)phonon interaction are considered. A linear model in which the phonon friction is taken to be linear in velocity of the hot electrons, and the other, in which the phonon friction is nonlinear in electron velocity. The nonlinearity considered here is a power-law type.
    In the following, we drive analytical expressions for the time-dependent non-equilibrium hot electron distribution for the two models of electron-(multi)phonon interaction.
\subsection{The phonon friction is proportional to the electron velocity V: linear model}
The kinetic equation for electron energy loss is:
\begin{equation}
\frac{{\partial}f_{e}(\epsilon,t)}{\partial t}+\dot{{\epsilon}}\frac{{\partial}f_{e}(\epsilon,t)}{\partial \epsilon} = - \frac{{\partial}f_{e}(\epsilon,t)}{\tau_r}.
\end{equation} 
                                                                              
Here,$ \dot{\epsilon} = (m\dot{v}^2)/2 = -\gamma_p v^2 = -{\epsilon}/{\tau_p}$ , where $\tau_r$ and $\tau_p$ are the radiative(photon) and the non-radiative(phonon) relaxation times, and $ \tau = t/{\tau_P}$ and $\alpha_{r-p} = \tau_r/\tau_p$  are dimensionless variables/parameters. With this, equation (1)  reduces  to,
\begin{equation}
\frac{{\partial}f_{e}(\epsilon,\tau)}{\partial {\tau}}-\frac{{\partial}f_{e}
(\epsilon,\tau)}{\partial ln\epsilon} = -\frac{{\partial}f_{e}(\epsilon,\tau)}
{\alpha_{r-p}}.
\end{equation}                      
Equation (2) can be solved analytically to give:
\begin{equation}
f_{e}(\epsilon,\tau) = f_0e^{-\tau/\alpha_{r-p}}\Theta(-\tau - ln[\epsilon/(\epsilon_F +\epsilon_L)])
\end{equation}   
                                                                                                                                                      
Here $\Theta[...]$ is a step function, and we have assumed an initial delta-function laser pulse of photon energy $\epsilon_L$ that excites electrons in the energy interval $\epsilon_F\leq\epsilon\leq\epsilon_F + \epsilon_L$. Inasmuch as Eq.(2) has only the forward propagating solutions (in energy space) we can readily incorporate the  slow process effective at the bottom of the conduction band by simply introducing a longer relaxation time $\tau_{pe} > \tau_{p}$  there. This gives rise to a kind of peaking effect, beacause of a pile-up of the hot electrons just above the Fermi energy of the degenerate electron subsystem. This peaking is due to the fermionic blocking of the electronic transitions, and it restricts the interaction phase space. 
Equation (3) is our basic result. In terms of it, we can calculate the total number $N_{pe}(\tau)$  of the hot electrons piled up near the Fermi energy, and also the total number of hot electrons $N_{hot}(\tau)$ above $\epsilon_F$. The number of hot electrons in the pile-up, $N_{>}(\tau)\equiv N_{hot}(\tau)-N_{pe}(\tau)$, with energy $\epsilon > \epsilon_F$ (for uniform excitation of scale height, $f_0$) is:
\begin{equation}
\frac{{\partial}N_{pe}(\tau)}{\partial{\tau}}+\frac{{\partial}N_{>}(\tau)}{\partial\tau} 
= -\frac{{\partial}N_{pe}(\tau)}{\beta_{pe}},\;\;\beta_{pe} = \frac{\tau_{pe}}{\tau_{p}}
\end{equation}
Now $N_>(\tau)=\int_{\epsilon_F}^{\epsilon_F+\epsilon_L}f_e(\epsilon,\tau)d\epsilon = f_0 e^{-\tau/\alpha_{r-p}}\int_{\epsilon_F}^{\epsilon_F\epsilon_L}\Theta(-\tau-ln(\epsilon/[\epsilon_F+\epsilon_L])d\epsilon$. 
We consider energy loss through phononic friction only, with this we get
\begin{equation}
N_{>}(\tau) = f_0(\epsilon_c+\epsilon_L)e^{-\tau}
\end{equation}
From equations (3) and (4) we have
\begin{equation}
\frac{{\partial}N_{pe}(\tau)}{\partial{\tau}}-f_0(\epsilon_F+\epsilon_L)e^{-\tau} = -\eta N_{pe}(\tau)\;\;,\;\;\eta = \frac{1}{\beta_{pe}}
\end{equation}
Solving by Laplace transforms we have
\begin{equation}
N_{pe}(\tau) = \frac{f_0(\epsilon_F + \epsilon_L)e^{-\eta \tau}}{1-\eta}[1-e^{-(1-\eta)\tau}],
\end{equation}
and the total number of hot electrons
\begin{equation}
N_{hot}(\tau) = N_{pe}(\tau) + N_>(\tau) = f_0(\epsilon_F + \epsilon_L)e^{-\tau}\left[1+ \frac{e^{(1-\eta)\tau}}{1-\eta}(1-e^{-(1-\eta)\tau})\right].
\end{equation}  
                                                                     
Time evolution of these two populations are plotted in Figs.(1) and (2).

\begin{figure}
\includegraphics[height=5.0cm,width=15.0cm]{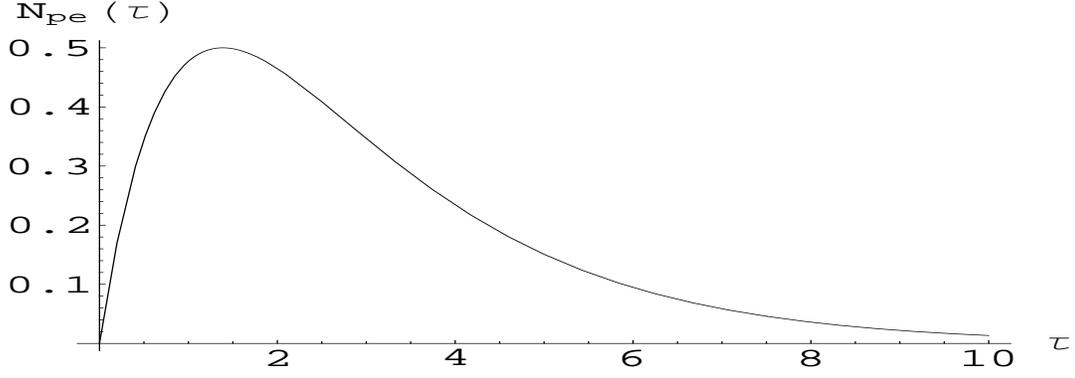}
\caption{Pile-up of hot electrons near the Fermi energy($\eta = 0.5$).}
\end{figure}
\begin{figure}
\includegraphics[height=5.0cm,width=15.0cm]{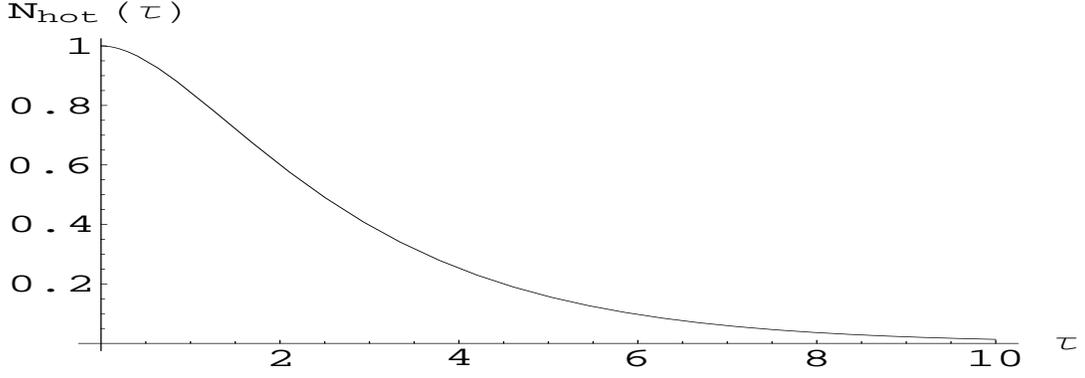}
\caption{Decay of hot electrons, $N_{hot}(\tau)$, for $\eta = 0.5$.}
\end{figure}
\subsection{The phonon friction is nonlinear (algebraic) in the electron velocity: nonlinear model}
\begin{equation}
\frac{{\partial}f_{e}(\epsilon,t;n)}{\partial t}+\dot{{\epsilon}}\frac{{\partial}f_{e}(\epsilon,t;n)}{\partial \epsilon} = - \frac{f_{e}(\epsilon,t;n)}{\tau_r}.
\end{equation}
With $\dot{\epsilon} = -\frac{\epsilon}{\tau_p}[\frac{\epsilon}{\epsilon_0}]^n$ and $\epsilon_0$ is an associated energy scale.

In line with the linear model, this kinetic equation has the solution,
\begin{equation}
f_{e}(\epsilon,\tau;n) = e^{-\tau/\alpha_{r-p}}\Theta\left(-\tau - \frac{1}{n}\left(\frac{\epsilon_0}{\epsilon_F + \epsilon_L}\right)^n\left[1-\left(\frac{\epsilon_F +\epsilon_L}{\epsilon}\right)^n\right]\right).
\end{equation}                  
The number of hot electrons $N_{>;n}(\tau)$ with energy greater than 
$\epsilon_F$ is 
\begin{equation}
N_{>;n}(\tau) = \int_{\epsilon_F}^{\epsilon_F + \epsilon_L}f_e(\epsilon,\tau;n)d\epsilon = f_{0;n}(\epsilon_F + \epsilon_L)\left[1+\frac{\tau}{\beta}\right]^{-1/n}.
\end{equation}
With $\beta = \frac{1}{n}(\chi)^n , \chi = \frac{\epsilon_0}{\epsilon_F + \epsilon_L}$.
As in the linear model, the number of hot electrons in the pile-up near the Fermi energy comes out to be:
\begin{equation}
N_{pe;n}(\tau) = f_0\epsilon_0\left[\frac{1}{n}\right]^{\frac{n+1}{n}}
\eta^{1/n}e^{-\eta(\tau + {\chi}^n/n)}\left[\int_{\eta\beta}^{\eta(\tau + \beta)}y^{-{\frac{n+1}{n}}}e^y dy \right].
\end{equation}
The total number of  hot electrons $"N_{hot;n}(\tau)"$ is $N_{hot;n}(\tau)
 = N_{>;n}(\tau) + N_{F;n}(\tau)$.
In the limit $\tau<<\beta$, or $ t << \frac{\tau_p}{n}\left(\frac{\epsilon_0}
{\epsilon_F + \epsilon_L}\right)^n $. The expression for hot electrons in the pile up, and total number of hot electrons will reduce to:
\begin{equation}
N_{pe;n}(\tau) = \frac{f_0(\epsilon_F + \epsilon_L)\chi^{-n}e^{-\eta\tau}}{\chi^{-n}-\eta}\left[1 - e^{-(\chi^{-n}-\eta)\tau}\right],
\end{equation}
\begin{equation}
N_{hot;n}(\tau) = f_0(\epsilon_F + \epsilon_L)\left[(1 + n \chi^{-n}\tau)^{-1/n}  + \frac{\chi^{-n}e^{-\eta\tau}}{\chi^{-n}-\eta}\left(1 - e^{-(\chi^{-n}-\eta)\tau}\right)\right].
\end{equation}
In the limit $n$ going to zero, we recover the results of the linear model.
\section{Incorporating pump-pulse duration}
Here, we consider the system  being pumped by a rectangular femtosecond pulse of  duration $t_p$. The effect of the pulse can be taken as the convolution integral of the rectangular pulse with the respective hot-electron  time-evolution curves(for the case n=0):
\begin{equation}
N_{f}^{p}(\tau) = \frac{1}{\tau_p}\int_{0}^{min(\tau_p,\tau)}N_{pe}(\tau-x)dx,
\end{equation}
\begin{equation}
N_{f}^{p}(\tau) = \frac{1}{\tau_p}\int_{0}^{min(\tau_p,\tau)}N_{hot}(\tau-x)dx.
\end{equation}
\begin{figure}
\includegraphics[height=15.0cm,width=7.0cm, angle=-90]{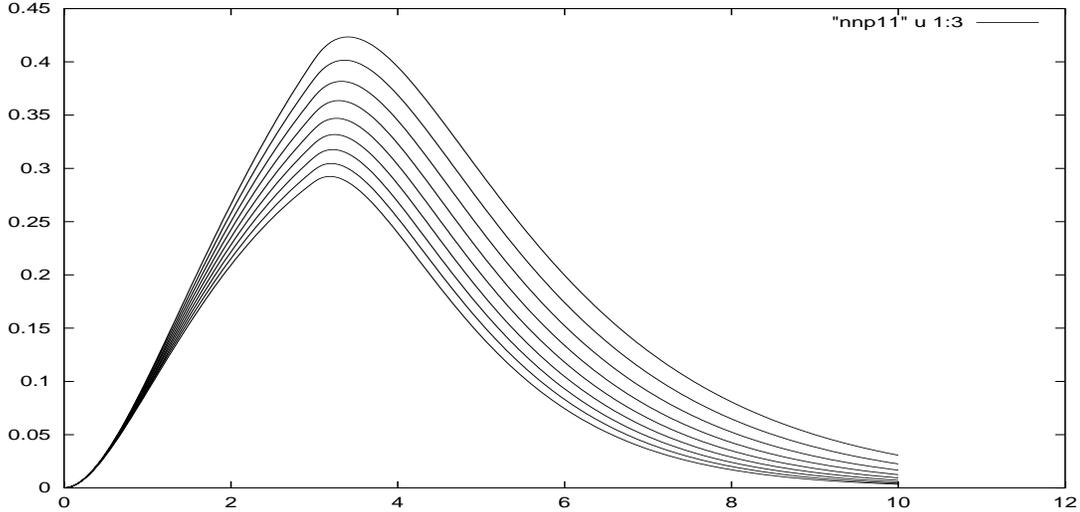}
\caption{Decay of hot electrons $N_{pe;p}(\tau)$ in the pile up as a function of $\tau$ in the presence of a rectangular laser pulse with width $\tau_p = 3$. Top most curve is for $\eta = 0$, lowest for $\eta = 0.9$ with a step of $0.1$}
\end{figure}
\begin{figure}
\includegraphics[height=15.0cm,width=7.0cm, angle= -90]{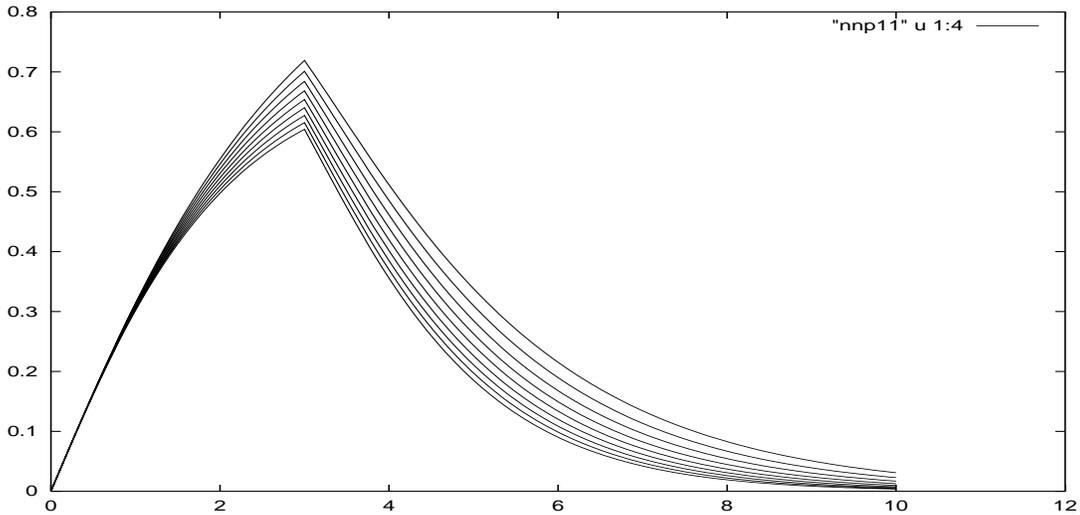}
\caption{Decay of hot electrons $N_{hot;p}(\tau)$ as a function of $\tau$,
with same set of parameters as in Figure 3}
\end{figure}
\section{Conclusion}
The distinctive feature of our calculated time evolution of the photoexcited
non-equilibrium electron distribution is its peaking at the Fermi energy, as readily seen in Figure 1. It reflects the fermionic blocking effect. This can be, and should be, probed in a femtosecond pump-probe experiment. This calculation refers to a situation not describable by the usual two-temperature model. Here, instead, the electron system itself is
comprised of two subsystems, one degenerate and the other nondegenerate. The
nondegenerate subsystem relaxes towards the degenerate subsystem by energy loss to the lattice.
\section{acknowledgement}
The author would like to thank Prof. N. Kumar for discussion and constant support.
\maketitle

\end{document}